\documentclass[prl,a4paper,twocolumn,noshowpacs,floatfix,groupedaddress,amsmath,amsfonts,amssymb,preprintnumbers,nofootinbib,citeautoscript]{revtex4}
\pdfoutput=1
\usepackage[T1]{fontenc}\usepackage[latin1]{inputenc}
\usepackage{dcolumn,graphicx,color,booktabs,wrapfig,microtype,afterpage} \graphicspath{{Figures/}}
\usepackage[charter,greekuppercase=italicized]{mathdesign}\usepackage{sidecap}
\renewcommand{\figurename}{\bf\textsf{Figure}\sffamily}
\renewcommand{\tablename}{Table}
\makeatletter\renewcommand{\fnum@figure}[1]{\textbf{\sffamily\figurename~\thefigure~|\,}}\makeatother
\makeatletter\renewcommand{\fnum@table}[1]{\textbf{\sffamily\tablename~\thetable~|\,}}\makeatother

\newcommand{\half}{\frac{1}{\protect\raisebox{0.7pt}{$\scriptstyle 2$}}}
\newcommand{\quarter}{\frac{1}{\protect\raisebox{0.7pt}{$\scriptstyle 4$}}}
\newcommand{\threequarters}{\frac{3}{\protect\raisebox{0.7pt}{$\scriptstyle 4$}}}
\newcommand{\eighth}{\frac{1}{\protect\raisebox{0.7pt}{$\scriptstyle 8$}}}

\newcount\hh \newcount\mm
\hh=\time \divide\hh by 60
\mm=\hh \multiply\mm by 60 \mm=-\mm
\advance\mm by \time
\def\now{\number\hh:\ifnum\mm<10{}0\fi\number\mm}

\definecolor{NatureBlue}{rgb}{0.012,0.3,0.63}
\usepackage[colorlinks,plainpages=false,linkcolor=NatureBlue,urlcolor=NatureBlue,citecolor=NatureBlue,pdfpagemode=UseNone,pdfstartview=FitBH]{hyperref}

\makeatletter
\newcommand{\bibstyle@supplement}{\bibpunct[, ]{[S}{]}{;}{n}{,}{,}%
    \gdef\bibnumfmt##1{[S##1]}}
\makeatother

\begin{document}


\title{\flushleft\fontsize{22pt}{26pt}\selectfont\sffamily\bfseries\textcolor{NatureBlue}{Intense low-energy ferromagnetic fluctuations in the antiferromagnetic heavy-fermion metal CeB$_6$}}

\author{\large\sffamily H.\,Jang$^{1}\!$} \author{\large\sffamily G.~Friemel$^{1}\!$} \author{J.~Ollivier$^2\!$} \author{A.\,V.~Dukhnenko$^3\!$} \author{N.\,Yu.\,Shitsevalova$^3\!$} \author{V.\,B.\,Filipov$^3\!$} \author{$\phantom{\quad\quad\kern5.2pt}$B.\,Keimer$^1\!$} \author{D.\,S.\,Inosov$^{1,4,\dagger}$\hfill\smallskip}
\affiliation{\flushleft\mbox{\sffamily $^1$\hspace{0.5pt}Max-Planck-Institut für Festkörperforschung, Heisenbergstraße 1, 70569 Stuttgart, Germany.}\\
\mbox{\sffamily $^2$\hspace{0.5pt}Institut Laue-Langevin, 6 rue Jules Horowitz, BP 156, 38042 Grenoble Cedex, France.}\\
\mbox{\sffamily $^3$\hspace{0.5pt}I.\,M.\,Frantsevich Institute for Problems of Material Sciences of NAS, 3 Krzhyzhanovsky str., 03680 Kiev, Ukraine.}\\
\mbox{\sffamily $^4$\hspace{0.5pt}Institut für Festkörperphysik, TU Dresden, D-01069 Dresden, Germany.}\\
\mbox{\sffamily $^\dagger$\hspace{0.5pt}E-mail:\,\href{mailto:Dmytro.Inosov@tu-dresden.de}{Dmytro.Inosov@tu-dresden.de}.}
}

\begin{abstract}\citestyle{nature}
\fontsize{8.9pt}{11pt}\selectfont
\noindent\textbf{Heavy-fermion metals exhibit a plethora of low-temperature ordering phenomena,\!\cite{SaxenaAgarwal00, SteppkeKuchler13, SchroderAeppli00, ChandraColeman02} among them the so-called hidden-order phases \cite{ChandraColeman02, SantiniCarretta09, KuramotoKusunose09, MydoshOppeneer11, OnoNakano13} that in contrast to conventional magnetic order are invisible to standard neutron diffraction. One of the oldest and \mbox{structurally simplest hidden-order compounds, CeB$_6$,} became famous for an elusive phase that was attributed to the antiferroquadrupolar ordering of cerium-4\textit{f} moments.\cite{Effantin85, NakaoMagishi01, MatsumuraYonemura09, MatsumuraYonemura12} In its ground state, CeB$_6$ also develops a more usual antiferromagnetic (AFM) order.\cite{Effantin85} Hence, its essential low-temperature physics was always considered to be solely governed by AFM interactions between the dipolar and multipolar Ce moments.\!\cite{ShiinaShiba97, ThalmeierShiina03} Here we overturn this estab\-lished perspective by uncovering an intense ferromagnetic (FM) low-energy collective mode that domi\-nates the magnetic excitation spectrum of CeB$_6$. Our inelastic neutron-scattering data reveal that the intensity of this FM excitation by far exceeds that of conventional spin-wave magnons emanating from the AFM wave vectors, thus placing CeB$_6$ much closer to a FM instability than could be anticipated. This propensity of CeB$_6$ to ferromagnetism may account for much of its unexplained behavior, such as the existence of a pronounced electron spin resonance,\!\cite{DemishevSemeno06, DemishevSemeno09} and should lead to a substantial revision of existing theories that have so far largely neglected the role of FM interactions.}
\end{abstract}

\pacs{78.70.Nx 75.30.Mb 75.30.-m 75.25.Dk\vspace{-0.7em}}

\pagestyle{plain}
\makeatletter
\renewcommand{\@oddfoot}{\hfill\bf\scriptsize\textsf{\thepage}}
\renewcommand{\@evenfoot}{\bf\scriptsize\textsf{\thepage}\hfill}
\makeatother

\citestyle{nature}
\maketitle\enlargethispage{0.6em}

\makeatletter\immediate\write\@auxout{\string\bibstyle{my-nature}}\makeatother
\renewcommand\bibsection{\section*{\sffamily\bfseries\footnotesize References\vspace{-10pt}\hfill~}}

\noindent

It was conjectured \cite{KrellnerForster08} that ferromagnetic (FM) correlations are a necessary precondition for the observability of an electron spin resonance (ESR) in Kondo-lattice systems. This general principle holds not only for compounds with static FM order, like CeRuPO or YbRh,\cite{KrellnerForster08, ForsterSichelschmidt10} but also for those with strong low-energy FM fluctuations arising from their proximity to a FM critical point, such as YbRh$_2$Si$_2$ \cite{SichelschmidtIvanshin03, SchaufussKataev09} or body-centered ($I$-type) YbIr$_2$Si$_2$. \cite{SichelschmidtWykhoff07} In the latter, direct and unambiguous demonstration of such fluctuations is rarely straightforward. For example, in weakly antiferromagnetic YbRh$_2$Si$_2$ their presence remained controversial until a recent inelastic-neutron-scattering (INS) study showed that a FM resonant mode can be stabilized at low temperatures by the application of an external magnetic field.\cite{StockBroholm12}

The heavy-fermion metal CeB$_6$ is most famous for an archetypal magnetically hidden order, the so-called antiferro-quadrupolar (AFQ) phase,\cite{SantiniCarretta09, KuramotoKusunose09} which sets in at low temperatures below $T_{\rm Q}=3.2$\,K and is usually associated with the ordering of magnetic quadrupole moments with the characteristic wave vector $\mathbf{Q}_{\rm AFQ}=R(\half \half \half)$ at the corner of the simple cubic Brillouin zone. This phase remains invisible to conventional neutron diffraction \cite{Effantin85} and therefore can be visualized only by resonant x-ray scattering or related probes that are directly sensitive to orbital degrees of freedom.\cite{NakaoMagishi01, MatsumuraYonemura09, MatsumuraYonemura12} Furthermore, conventional AFM order with a double-$\mathbf{Q}$ commensurate magnetic structure, $\mathbf{Q}_{{\rm AFM}_1}\!=\!\Sigma(\quarter \quarter 0)$ and $\mathbf{Q}_{{\rm AFM}_2}\!=S(\quarter \quarter \half)$, is found below $T_{\rm N}=2.3$\,K. \cite{Effantin85, ZaharkoFischer03} An intimate interplay between the AFQ and AFM order parameters is suggested by our recent observation of the magnetic resonant mode that forms at $\mathbf{Q}_{\rm AFQ}$ below $T_{\rm N}$ in close resemblance to that found in unconventional superconductors.\cite{FriemelLi12, AkbariThalmeier12}

It has been generally acknowledged that the ordering phenomena in CeB$_6$ must be governed by AFM interactions between multipolar moments of the Ce-4$f$ electrons mediated by the itinerant conduction electrons.\cite{ShiinaShiba97, ThalmeierShiina03} At the same time, CeB$_6$ is also known for its clearly observable narrow ESR signal \cite{DemishevSemeno06, DemishevSemeno09} that is rather suggestive of FM correlations. \cite{Schlottmann12} Yet they have, up to now, escaped any direct observation \cite{ZaharkoFischer03} and remained largely neglected in theoretical models. Such an apparent discord between the results of different experimental probes calls for a more detailed reconsideration of magnetic interactions in this model system.

\begin{figure}[t]\vspace{-0.35em}
\includegraphics[width=\columnwidth]{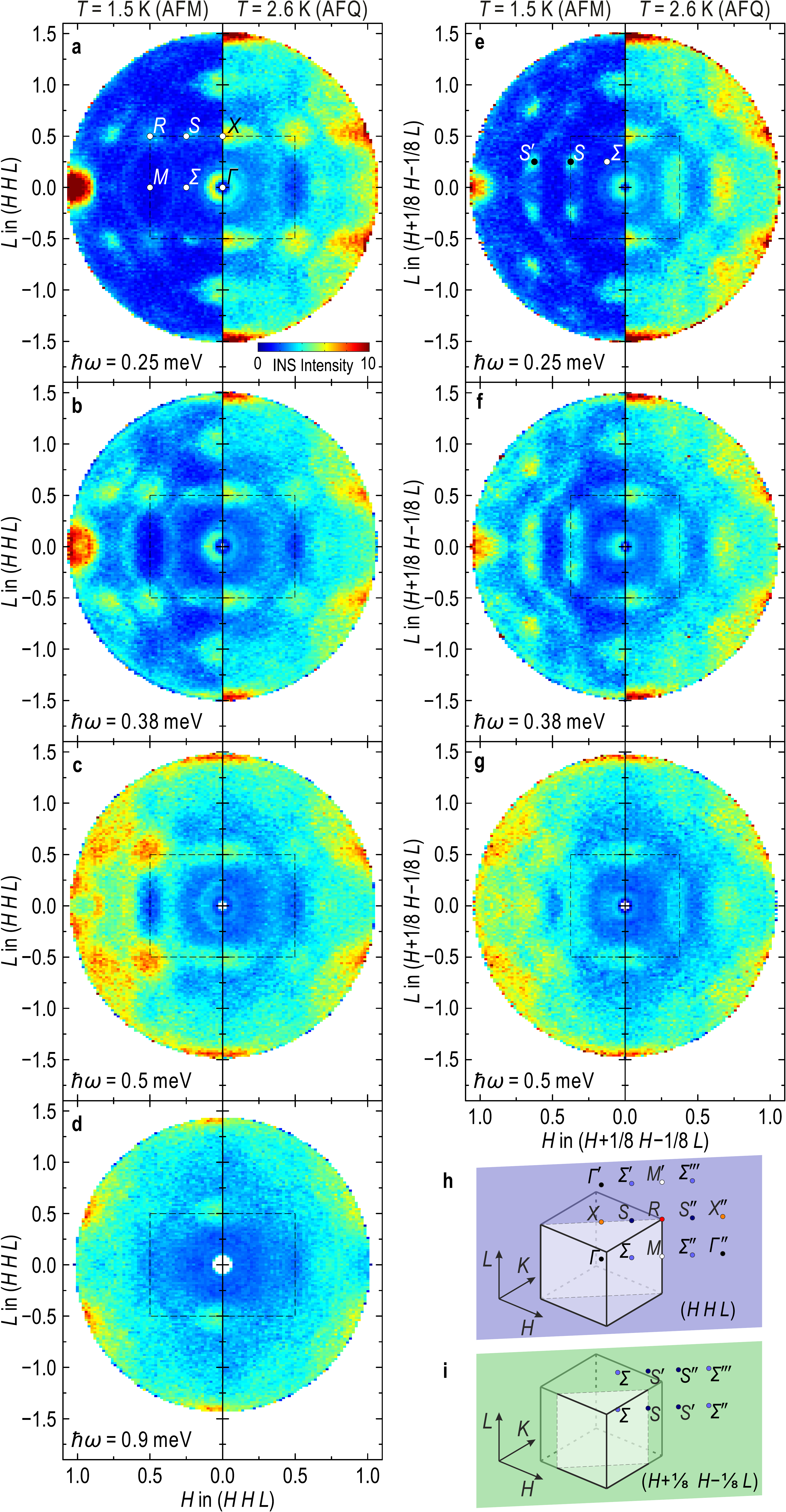}
\caption{\textbf{Constant-energy maps obtained from the TOF data.} \textbf{a}\,--\,\textbf{d}, $(H\,H\,L)$ cuts at $\hslash\omega = 0.25$, 0.38, 0.5, and 0.9~meV. \textbf{e}\,--\,\textbf{g}, $(H\!+\!\eighth~H\!-\!\eighth~L)$ cuts at $\hslash\omega = 0.25$, 0.38, and 0.5~meV. The data were symmetrized about the natural mirror planes of the reciprocal space. The left and right halves of every panel were measured in the AFM and AFQ states, respectively. The integration range in energy was $\pm$0.1\,meV around the average values displayed in the panels. \textbf{h},\textbf{i}, Location of the $(H\,H\,L)$ and $(H\!+\!\eighth~H\!-\!\eighth~L)$ planes in the reciprocal space with respect to the cubic Brillouin zone of CeB$_6$ and the labeling of high-symmetry points.}
\label{Fig:Qmaps}
\end{figure}

To resolve this controversy, we have carried out a high-resolution (${\scriptstyle\Delta}E\approx80$\,$\mu$eV) time-of-flight (TOF) neutron-scattering experiment on a large single crystal of CeB$_6$ (see Methods) to obtain the complete momentum-resolved spectrum of magnetic excitations in the low-energy range, $\hslash\omega \leq 2$\,meV, over the whole volume of the Brillouin zone. The measurements were done at two temperatures, $T=1.5$\,K and 2.6\,K, i.e. in the AFM and AFQ states, respectively.

In Fig.\,\ref{Fig:Qmaps}, we present constant-energy cuts taken from our 4-dimensional datasets along the two parallel horizontal planes, $(H\,H\,L)$ and $(H\!+\!\eighth~H\!-\!\eighth~L)$, as clarified in Fig.\,\ref{Fig:Qmaps}h,i. Cuts along other directions are shown in Fig.\,\ref{Fig:S1} of the Supplementary Information (SI). In every panel, we compare equivalent datasets acquired at $T=1.5$\,K (left) and 2.6\,K (right), whereas different rows of panels represent several energy ranges, integrated within $\pm$0.1\,meV around the displayed average values. The strongest inelastic intensity can be observed in the AFM state at 0.25\,meV (Fig.\,\ref{Fig:Qmaps}a) near the zone center, $\Gamma(110)$. Its strong temperature dependence proves that it cannot originate from scattering on the acoustic phonon modes, but must represent a previously unknown magnetic excitation centered at the FM wave vector. The same mode, but with weaker intensity, can also be seen at the equivalent $\Gamma(001)$ point. This enhancement of intensity towards longer scattering vectors can result from an anomalous non-monotonic $Q$-dependence of the magnetic form factor, which is typical for higher-rank multipole order parameters.\cite{KuramotoKusunose09} At higher energies (Fig.\,\ref{Fig:Qmaps}b,c), the intense feature at the $\Gamma$ point broadens and forms a ring around the zone center, characteristic of a dispersive soft mode.

Concurrently, at 0.5\,meV (Fig.\,\ref{Fig:Qmaps}c) we observe the previously reported\cite{FriemelLi12, AkbariThalmeier12} resonant excitation centered at the AFQ wave vector, $R(\half \half \half)$. At 0.9\,meV (Fig.\,\ref{Fig:Qmaps}d), a broader exciton mode with an anisotropic shape, elongated in the (110) direction, appears at the $X(0 0 \smash{\half})$ point. In addition, Fig.\,\ref{Fig:Qmaps}a,e shows notably weaker intensity spots near the AFM ordering vectors, such as $S(\quarter \quarter \half)$, $S^\prime(\threequarters \half \quarter)$, $S^{\prime\prime}(\threequarters \threequarters \half)$ or equivalent. They originate from conventional spin-wave modes that either vanish or broaden considerably above $T_{\rm N}$.

\begin{figure}[t!]\vspace{-1pt}
\includegraphics[width=0.95\columnwidth]{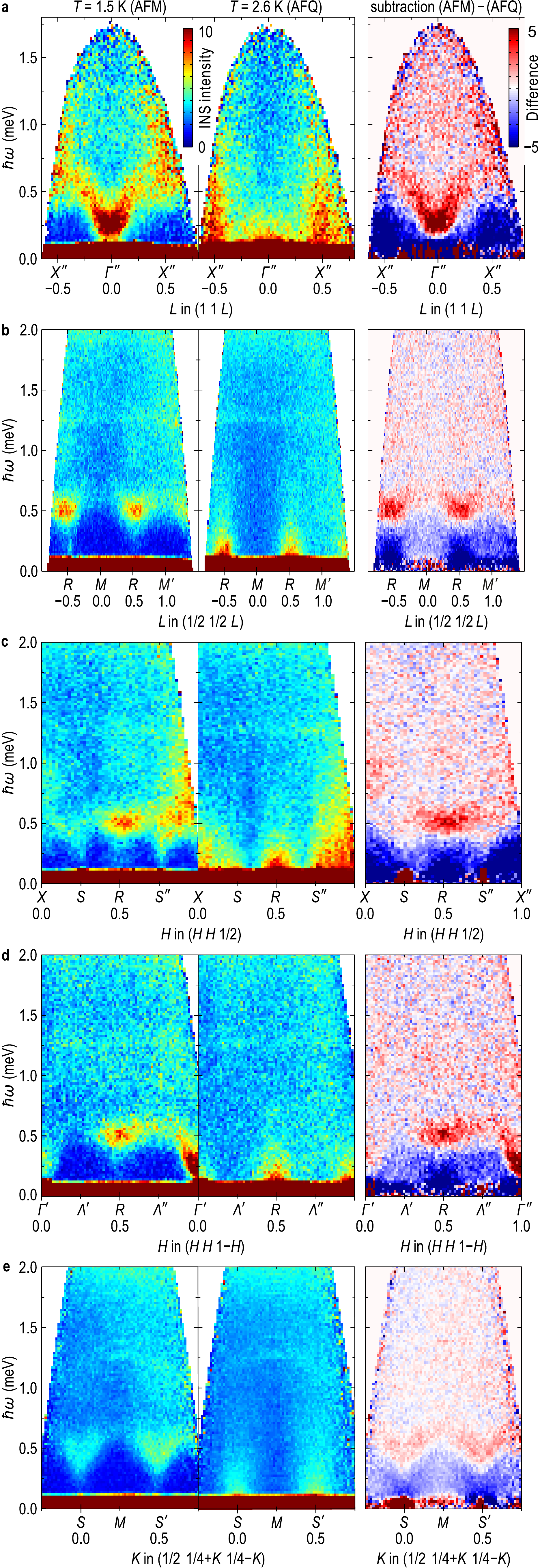}
\caption{\textbf{Energy-momentum cuts.} In each panel, the first two images show unprocessed data at 1.5\,K (AFM state) and 2.6\,K (AFQ state), whereas the rightmost image shows the corresponding intensity difference along the following high-symmetry directions: \textbf{a},~$(1\,1\,L)$; \textbf{b},~$(\half\,\half\,L)$; \textbf{c},~$(H\,H\,\half)$; \textbf{d},~$(H\,H\,1\!-\!H)$; \textbf{e},~$(\smash{\half}~\smash{\quarter}\!+\!K~\smash{\quarter}\!-\!K)$.\vspace{-3em}}
\label{Fig:QEmaps}
\end{figure}

To get an insight into the dispersion and interrelation of these numerous excitations, in Fig.\,\ref{Fig:QEmaps} we show several energy-momentum cuts along high-symmetry directions, measured in the AFM and AFQ states, together with their respective intensity differences. The latter illustrate the redistribution of spectral weight across $T_{\rm N}$, as the collective modes (red) are destroyed above the transition, and their intensity fills in the spin-gap region (blue) at lower energies. The $(1\,1\,L)$ cut in Fig.\,\ref{Fig:QEmaps}a crosses the intense FM excitation at the $(110)$ wave vector. One can clearly see a narrow dispersive mode in the AFM state, centered at 0.25\,meV above a spin gap of $\sim$\,0.2\,meV. Away from the $\Gamma$ point, this mode disperses towards higher energies, reaching above 0.5\,meV at the zone boundary. The mode is destroyed above $T_{\rm N}$, giving rise to an intense quasielastic signal centered at zero energy.

A parallel cut along the $(\half\,\half\,L)$ line that crosses two equivalent $R$ points is shown in Fig.\,\ref{Fig:QEmaps}b. Here, two resonant peaks at 0.5\,meV can be clearly seen.\cite{FriemelLi12} They reveal only a weak upward dispersion of no more than 0.2\,meV between the $R$ and $M$ points. It is also illustrative to consider cuts along the $(H\,H\,\half)$ line (Fig.\,\ref{Fig:QEmaps}c) that connects the $R$ point to the AFM ordering vectors, $S$ and $S^{\prime\prime}\!\!$, and along the $(H\,H\,1\!-\!H)$ diagonal (Fig.\,\ref{Fig:QEmaps}d) connecting it to the two zone centers, $\Gamma^\prime(001)$ and $\Gamma^{\prime\prime}(110)$. In both cross-sections, the resonant mode at $R$ represents a distinct local maximum of intensity that is continuously linked by weaker branches of gapped excitations either to the spin-wave modes emanating from the $S$ points or to the FM mode at $\Gamma$. This justifies its consideration as a separate exciton mode, localized both in energy and in $\mathbf{Q}$-space. In Fig.\,\ref{Fig:QEmaps}d, the peak at $R$ appears as a swelling on the upward-dispersing branches emerging from the two $\Gamma$ points, which are softened towards lower energies near the zone boundary. This is reminiscent of the situation in the UPd$_2$Al$_3$ heavy-fermion superconductor, where the low-energy mode emanating from the magnetic zone center also forms an anomalous local minimum in the dispersion and a subsidiary maximum of intensity near the zone boundary.\cite{HiessBernhoeft06}

Next, we consider the spin-wave excitations, which are best revealed in Fig.\,\ref{Fig:QEmaps}e along the $(\half~\quarter\!+\!K~\quarter\!-\!K)$ cut passing through two equivalent AFM ordering vectors, $S(\smash{\half} \smash{\quarter} \smash{\quarter})$ and $S^\prime(\half \threequarters \frac{\overline{1}}{\protect\raisebox{0.7pt}{$\scriptstyle 2$}})$, as well as in Fig.\,\ref{Fig:QEmaps}c. The spin-wave modes emanating from these wave vectors are characterized by a spin gap of about 0.3--0.4\,meV and reach up to 0.7\,meV at the zone boundary ($M$ point), which results in a narrow total band width that is comparable to the spin-gap energy.

\begin{figure*}[t]\vspace*{-2.0em}
\includegraphics[width=\textwidth]{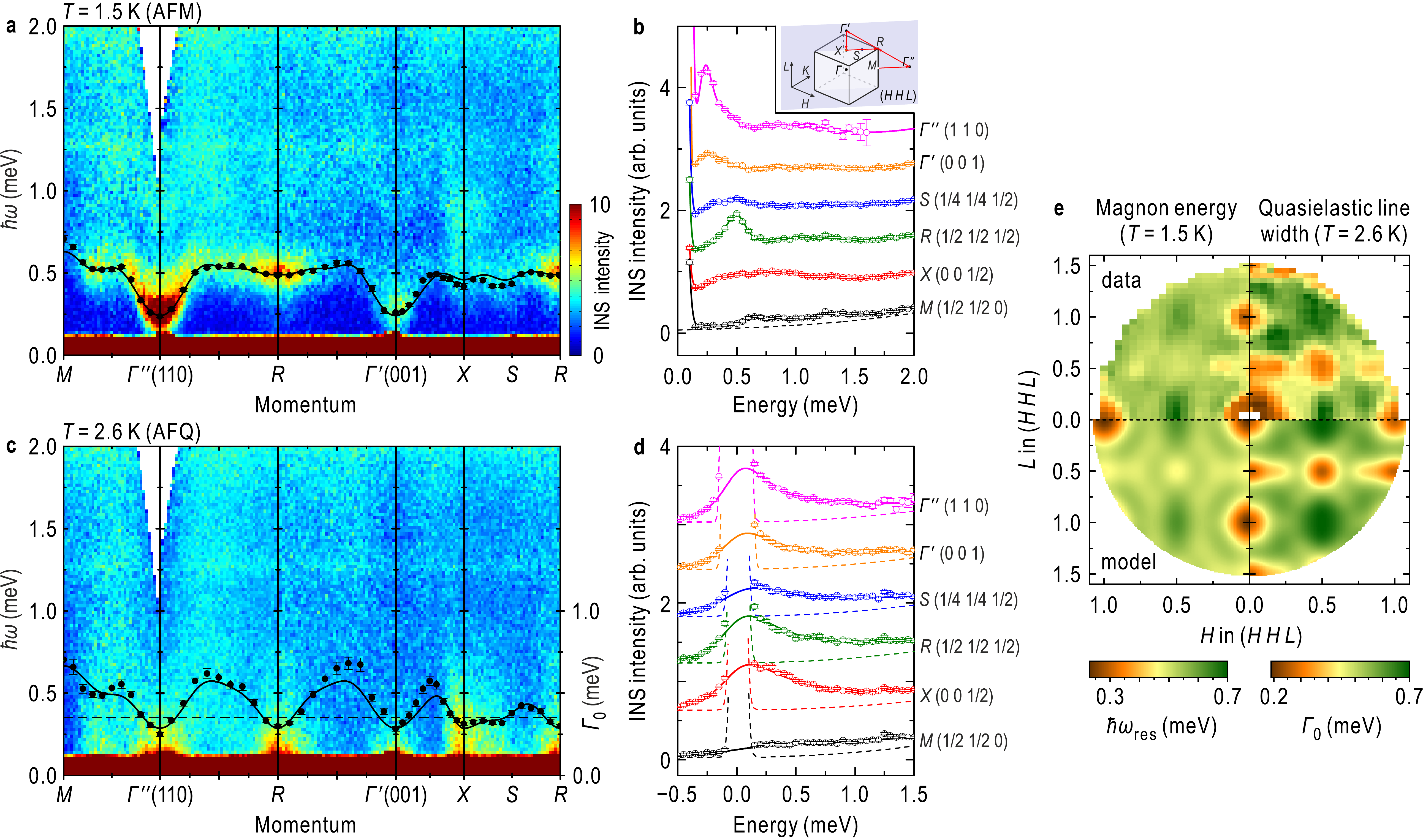}\vspace*{-0.5em}
\caption{\textbf{Correlation between the dispersion of collective mag\-non excitations in the AFM state and the momentum-dependent quasielastic line width in the AFQ state.} \textbf{a},~Energy-momentum profile along high-symmetry directions (see inset) at $T=1.5$\,K. Exciton energies, determined as peak maxima, are overlayed in black and fitted in accord with the cubic symmetry of the Brillouin zone. \textbf{b},~INS spectra at $T=1.5$\,K taken at several high-symmetry points. Solid lines are empirical fits to the data. \textbf{c},~Energy-momentum profile at $T=2.6$\,K along the same trajectory as in panel a. The quasielastic line width $\Gamma$ is plotted in black. The dashed line shows the powder-averaged line width from Ref.~\citenum{HornSteglich81}, taken for the same temperature. \textbf{d},~Quasielastic fits of the $T=2.6$\,K spectra at high-symmetry points. In panels b and d, the datasets were shifted vertically in increments of 0.6 units for clarity. The estimated background level is shown with dashed lines. \textbf{e},~Magnon energy (left) and quasielastic line width (right) presented as color maps within the $(HHL)$ plane. The value of every pixel in the upper half of the panel was obtained by fitting the corresponding energy-dependence curve, whereas the lower half depicts an empirical model following the Brillouin-zone symmetry, fitted to the same data (see Supplementary Information).\vspace{-0.3em}}
\label{Fig:Dispersion}
\end{figure*}

In total, we can distinguish three distinct types of low-energy magnetic excitations in CeB$_6$: a strong FM soft mode at the zone center, a subsidiary maximum of intensity at the $R$ point, and the spin-wave modes emanating from the AFM wave vectors. All these excitations hybridize to form a continuous dispersive magnon band in a narrow energy range between 0.2 and 0.7\,meV. This is illustrated in Fig.\,\ref{Fig:Dispersion}a, where we plot the intensity distribution along main high-symmetry directions of the reciprocal space with the fitted dispersion relation. An extended version of this dataset is given in Fig.\,\ref{Fig:S2} of the SI. At higher energies, one can discern additional magnetic intensity near the $\Gamma$, $R$ and $X$ points, which appears as broad secondary maxima in the one-dimensional spectra presented in Fig.\,\ref{Fig:Dispersion}b. The magnetic origin of these higher-energy modes can be deduced from their temperature dependence, as they clearly vanish above $T_{\rm N}$ according to Fig.\,\ref{Fig:QEmaps} and therefore appear as positive intensity in the corresponding subtraction plots. However, due the broadness and weakness of these modes, their dispersion could not be followed continuously across the Brillouin zone with the available experimental statistics.

In contrast to the early theoretical predictions, \cite{ThalmeierShiina03} we find no dispersive magnon excitations in the AFQ state. Immediately above $T_{\rm N}$ (Fig.\,\ref{Fig:Dispersion}c), the magnetic intensity collapses into a broad quasielastic peak centered at zero energy, which can be well described by the conventional quasielastic Lorentzian line shape, $S(\mathbf{Q},\omega)\propto[n(\omega)+1]\,\hslash\omega\Gamma_0/(\hslash^2\omega^2+\Gamma_0^2)$, as shown in Fig.\,\ref{Fig:Dispersion}d for several high-symmetry points. Here, $\hslash\omega$ is the energy transfer and $n(\omega)$ is the Bose population factor. The temperature-dependent linewidth, $\Gamma_0$, was previously reported only for polycrystalline samples within the paramagnetic state, $T>T_{\rm Q}$, and was therefore assumed to be momentum-independent.\cite{HornSteglich81} Remarkably, our analysis reveals a strong $\mathbf{Q}$-dependence of the quasielastic line width within the AFQ state, as plotted in Fig.\,\ref{Fig:Dispersion}c with black circles. Its minima are found at the $\Gamma$, $R$ and $X$ points, where intense magnetic excitations form below $T_{\rm N}$. Moreover, the juxtaposition of Figs.\,\ref{Fig:Dispersion}a and \ref{Fig:Dispersion}c suggests a direct correlation between $\Gamma_0$ and the magnon energy in the AFM state, $\hslash\omega_{\rm res}$, implying that the strongly overdamped signal in the AFQ state carries the essential primordial information about the energies of the would-be collective modes already well above the AFM transition. This conclusion is further corroborated by Fig.\,\ref{Fig:Dispersion}e, where we directly compare $\hslash\omega_{\rm res}$ with $\Gamma_0$ within the $(HHL)$ plane of the reciprocal space. The quasielastic signal also has a rich structure in momentum space, as revealed by the lowest-energy intensity maps in Figs.\,\ref{Fig:Qmaps}a and\,\ref{Fig:Qmaps}e, measured in the AFQ state. In analogy to the hidden-order phase of URu$_2$Si$_2$,\!\cite{WiebeJanik07} numerous maxima of the scattering function in $\mathbf{Q}$-space can possibly originate from nesting vectors of the normal-state Fermi surface of CeB$_6$, which until now has not been measured directly. If so, this structure could hold a key to the microscopic origin of the complex magnon spectrum, which we revealed in the present study.

The experiments presented here have several important implications. First, the propensity of CeB$_6$ to a FM instability evidenced by the intense exciton mode at the $\Gamma$ point resolves the long-standing controversy about the role of FM correlations for the enigmatic low-temperature properties of this structurally simple material. It explains the appearance of sharp ESR lines observed in a broad range of magnetic fields \cite{DemishevSemeno06, DemishevSemeno09} and draws exciting new parallels between CeB$_6$ and a number of other well-known Kondo-lattice systems, such as YbRh$_2$Si$_2$ or YbIr$_2$Si$_2$, where the proximity to a FM quantum critical point was established. \cite{SichelschmidtIvanshin03, SchaufussKataev09, SichelschmidtWykhoff07} Second, our results are of model character for a much broader class of Ce-based heavy-fermion systems, such as the novel clathrate compound Ce$_3$Pd$_{20}$Si$_6$,\cite{CustersLorenzer12} whose phase diagram is qualitatively identical to that of CeB$_6$, exhibiting a magnetically hidden AFQ phase side-by-side with AFM ordering. \cite{OnoNakano13}

Finally, the vivid importance of FM correlations should lead to a substantial revision of the existing theories behind the ordering phenomena in CeB$_6$, which were so far relying exclusively on the AFM coupling between dipolar and multipolar moments of the Ce-4$f$ electrons. The complete high-resolution mapping of the energy-momentum space, which became possible in this work due to the recent advances in the cold-neutron TOF spectrometer instrumentation, offers a strict testing ground for such new theoretical models.

\vspace{0.8em}
{\footnotesize\noindent
{\sffamily\textbf{Methods.}}\vspace{1ex}\enlargethispage{2pt}

\noindent \textbf{Sample preparation and instrumental setup.} A single crystal of CeB$_6$ with a mass of 4\,g was grown by the floating-zone method from a 99.6\,at.\,\% isotope-enriched $^{11}$B powder, as described elsewhere. \cite{FriemelLi12} INS data were collected using the cold-neutron time-of-flight spectrometer IN5 at the high-flux reactor of the Institute Laue-Langevin (ILL), Grenoble, France. This instrument is equipped with a 30\,m$^2$ position-sensitive detector comprised of 10$^5$ pixels, covering 147$^\circ$ of azimuthal angle and $\pm$20$^\circ$ out of the equatorial plane. The sample was mounted into a standard cryostat with its (110) and (001) directions in the equatorial plane. The incident neutron wavelength was fixed at 5\,\AA\ (3.27 meV), yielding energy resolution (full width at half maximum) of 0.08\,meV at zero energy transfer. The measurements were taken by rotating the crystal about the vertical $(1\overline{1}0)$ axis, which were then combined and transformed into the energy-momentum space using the HORACE analysis software.\smallskip

\noindent \textbf{Momentum-space notation.} Throughout this paper, reciprocal-lattice vectors are indexed on the simple-cubic unit cell (space group $Pm3m$, lattice constant $a=4.1367$\,\AA). The wave-vector coordinates are given in reciprocal lattice units (1~r.l.u.~=~$2\pi/a$). In labeling the high-symmetry points of the Brillouin zone, we follow the conventional notation given in Fig.\,\ref{Fig:Qmaps}h,i.~\vspace{-0.5em}

\noindent\textbf{\sffamily Acknowledgements.}~We are grateful to S.\,V.~Demishev, V.~Kataev, J.~Sichel\-schmidt, and P.~Thalmeier for enlightening discussions.\smallskip

\noindent{\textbf{\sffamily Author~contributions.} A.V.\,D., N.\,Yu.\,S. and V.~B.\,F. synthesized the single-crystalline sample. H.\,J., G.\,F. and D.\,S.\,I. performed the INS experiments and analyzed the data. H.\,J., G.\,F., B.\,K. and D.\,S.\,I. developed the physical interpretation. J.\,O. provided instrument support at ILL. H.\,J. and D.\,S.\,I. designed the figures and wrote the manuscript. B.\,K. and D.\,S.\,I. supervised the project.}\smallskip

\noindent{\textbf{\sffamily Author information.} The authors declare no competing financial interests. Correspondence and requests for materials should be addressed to D.\,S.\,I. $\langle$\href{mailto:Dmytro.Inosov@tu-dresden.de}{Dmytro.Inosov@tu-dresden.de}$\rangle$.}\bigskip\vfill

\vspace{1em}
\vfill
}

\onecolumngrid\clearpage

\vfill
\clearpage

\renewcommand\thefigure{S\arabic{figure}}
\renewcommand\thetable{S\arabic{table}}
\renewcommand\theequation{S\arabic{equation}}
\renewcommand\bibsection{\section*{\sffamily\bfseries\footnotesize Supplementary References\vspace{-6pt}\hfill~}}

\citestyle{supplement}

\pagestyle{plain}
\makeatletter
\renewcommand{\@oddfoot}{\hfill\bf\scriptsize\textsf{S\thepage}}
\renewcommand{\@evenfoot}{\bf\scriptsize\textsf{S\thepage}\hfill}
\renewcommand{\@oddhead}{H.~Jang \textit{et~al.}\hfill\Large\textsf{\textcolor{NatureBlue}{SUPPLEMENTARY INFORMATION}}}
\renewcommand{\@evenhead}{H.~Jang \textit{et~al.}\Large\textsf{\textcolor{NatureBlue}{SUPPLEMENTARY INFORMATION}}\hfill}
\makeatother
\setcounter{page}{1}\setcounter{figure}{0}\setcounter{table}{0}\setcounter{equation}{0}

\makeatletter\immediate\write\@auxout{\string\bibstyle{my-apsrev}}\makeatother

\onecolumngrid\normalsize

\begin{center}{\vspace*{0.1pt}\Large{Supplementary Information to the Letter\smallskip\\\sl\textbf{``\hspace{1pt}Intense low-energy ferromagnetic fluctuations\\in the antiferromagnetic heavy-fermion metal CeB$_6$''}}}\end{center}\bigskip

\twocolumngrid

\begin{figure}[b]
\includegraphics[width=\columnwidth]{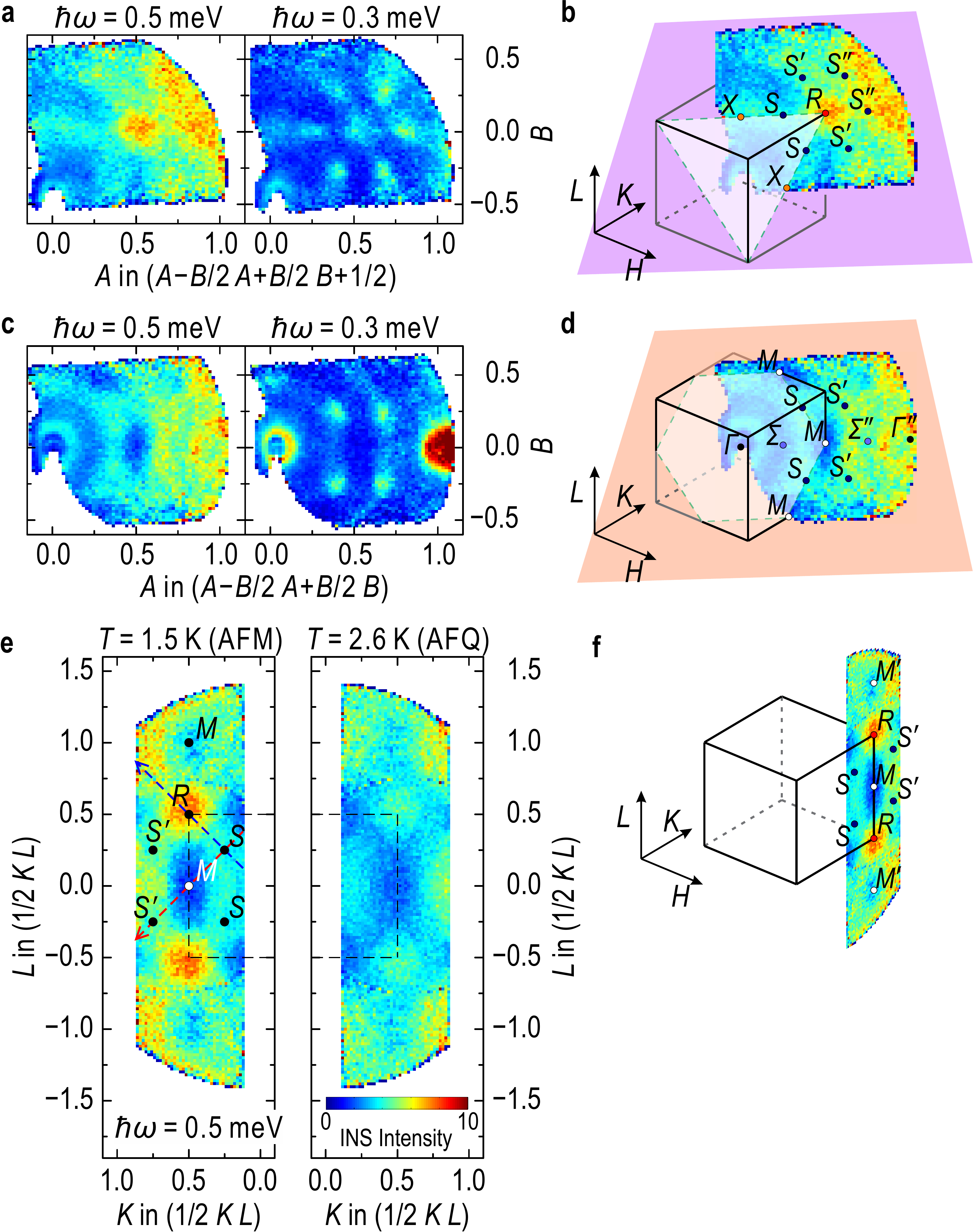}
\caption{\textbf{Additional constant-energy cuts along different high-symmetry planes.} \textbf{a},~Data at $\hslash\omega\!=\!0.5$\,meV (left) and 0.3\,meV (right) in the AFM state ($T\!=\!1.5$\,K) along the hexagonal $(A\!-\!\smash{\half}B~A\!+\!\smash{\half}B~B\!+\!\smash{\half})$ plane, perpendicular to the $(1\overline{1}1)$ direction and passing through the $R(\half\half\half)$ point, as shown in panel~\textbf{b}. \textbf{c},~The same for a parallel $(A\!-\!\smash{\half}B~A\!+\!\smash{\half}B~B)$ plane, passing through two $\Gamma$ points, $(000)$ and $(110)$, and multiple $S$ points, as shown in panel~\textbf{d}. \textbf{e},~Constant-energy cuts at $\hslash\omega\!=\!0.5$\,meV for $T\!=\!1.5$\,K (left) and $T\!=\!2.6$\,K (right) along the $(\half K L)$ plane (Brillouin-zone face), passing through a pair of $R$ points, as illustrated in panel \textbf{f}.\vspace{-1em}}
\label{Fig:S1}
\end{figure}\enlargethispage{1em}

\begin{figure*}[t]
\includegraphics[width=\textwidth]{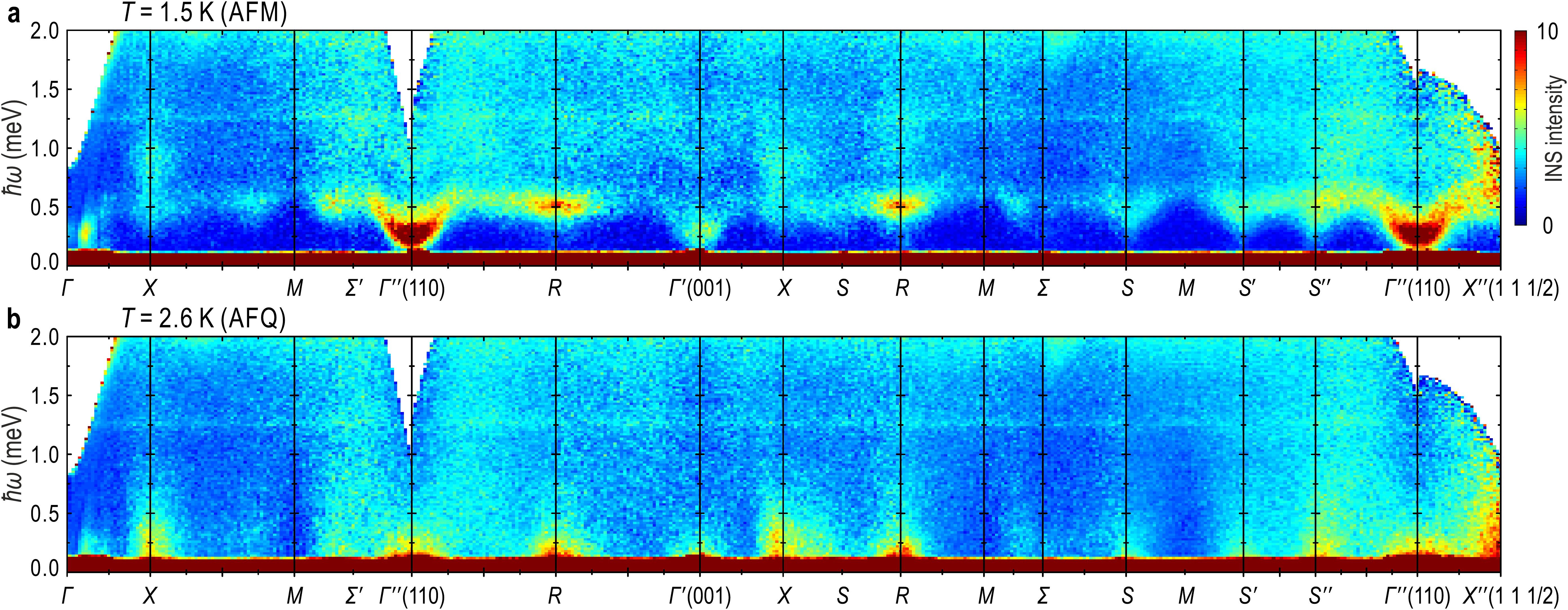}
\caption{\textbf{Extended energy-momentum profiles along a polygonal path through high-symmetry directions in $\mathbf{Q}$-space.} \textbf{a},~Collective modes in the AFM state ($T=1.5$\,K). \textbf{b},~Quasielastic intensity distribution in the AFQ state ($T=2.6$\,K). The leftmost $\Gamma$ point corresponds to the direct beam ($\mathbf{Q}=0$).\vspace{-0.5em}}
\label{Fig:S2}
\end{figure*}

\noindent\textbf{\it Constant-energy profiles.}\vspace*{5pt}

In Fig.\,\ref{Fig:S1}, we present additional INS data, obtained as two-dimensional cuts through the four-dimensional TOF dataset along different directions that are not parallel to the equatorial scattering plane. The cuts in Fig.\,\ref{Fig:S1}a were taken at the resonance energy ($\hslash\omega=0.5$\,meV) and below it ($\hslash\omega=0.3$\,meV) in the AFM state along the plane orthogonal to the $(1\overline{1}1)$ direction and passing through the $R$ point. The resonant mode can be seen in the center of the panel, surrounded by a hexagonal arrangement of spin-wave branches emanating from six equivalent $S$ points as clarified in Fig.\,\ref{Fig:S1}b. Two parallel cross-sections through the zone center, taken at the same energies, are shown in Fig.\,\ref{Fig:S1}c. Here, one sees the intense FM mode at low energies along with the spin-wave intensity appearing as sharp but much weaker intensity spots in a hexagonal arrangement. The location of the cut in the Brillouin zone is explained by Fig.\,\ref{Fig:S1}d.

The cross-section presented in Fig.\,\ref{Fig:S1}e shows data in both AFM and AFQ states along the face of the cubic Brillouin zone, which is orthogonal to the (100) direction, taken at the resonance energy ($\hslash\omega=0.5$\,meV). In contrast to the diagonal of the Brillouin zone (Fig.\,\ref{Fig:Qmaps}c), where the two resonant modes appear to be connected by weaker streaks of intensity forming a continuous elliptical feature around the $M$ point, here one can clearly see that the ``ellipse'' is disconnected from the resonance (this is best seen in the $T=2.6$\,K data). Its intensity is maximized in between the two $S$ points and is therefore likely stemming from the upper part of the spin-wave excitation branch that connects these points and lies in approximately the same energy range, as one can observe from the fit in Fig.\,\ref{Fig:Dispersion}a.

\vspace*{10pt}\noindent\textbf{\it Energy-momentum profiles.}\vspace*{5pt}

On the next page, in Fig.\,\ref{Fig:S2}, we also present an extended version of the energy-momentum profiles along all high-symmetry directions of the $\mathbf{Q}$-space, extracted from the $T=1.5$\,K (top) and $T=2.6$\,K (bottom) datasets. Here, both the dispersion of the collective magnon modes (AFM state) and the intensity modulation of the quasielastic response (AFQ state) can be clearly seen over the broad range of momenta. Note the existence of additional weak temperature-dependent intensity at higher energies ($\hslash\omega>0.8$\,meV), in particular near the $X$ and $\Gamma$ points.

\vspace*{10pt}\noindent\textbf{\it Fitting model.}\vspace*{5pt}

\begin{table}[b!]
\begin{center}
\begin{tabular}{lc@{~~}r@{~~}r}
    \toprule
        ~  &   $R_j/a$   & $\hslash\omega_{\rm res}$~~ & $\Gamma_0$~~~~~~\\
	\midrule
        $A_0$   &   (0,0,0)  & 0.50927   &   0.47760  \\
        $A_1$   &   (1,0,0)  & --$\kern.5pt$0.00771    &   --$\kern.5pt$0.00507  \\
        $A_2$   &   (1,1,0)  & --$\kern.5pt$0.00559    &   --$\kern.5pt$0.00348  \\
        $A_3$   &   (1,1,1)  & 0.00090    &   0.01570  \\
        $A_4$   &   (2,0,0)  & --$\kern.5pt$0.00223    &   --$\kern.5pt$0.00786  \\
        $A_5$   &   (2,1,0)  & --$\kern.5pt$0.00471    &   --$\kern.5pt$0.00637  \\
        $A_6$   &   (2,1,1)  & --$\kern.5pt$0.00377    &   --$\kern.5pt$0.00623  \\
        $A_8$   &   (2,2,0)  & 0.00112    &   0.00351  \\
        $A_{9a}$   &   (2,2,1)  & 0.00100    &  0.00097  \\
        $A_{9b}$   &   (3,0,0)  & --$\kern.5pt$0.00098    &   0.00347  \\
        $A_{10}$   &   (3,1,0)  & --$\kern.5pt$0.00016    &   0.00116  \\
        $A_{11}$   &   (3,1,1)  & 0.00108    &   0.00068 \\
        $A_{12}$   &   (2,2,2)  & --$\kern.5pt$0.00073    &   --$\kern.5pt$0.00269  \\
        $A_{13}$   &   (3,2,0)  & --$\kern.5pt$0.00065    &   --$\kern.5pt$0.00015  \\
        $A_{14}$   &   (3,2,1)  & 0.00056    &   0.00047  \\
        $A_{16}$   &   (4,0,0)  & --$\kern.5pt$0.00096    &   --$\kern.5pt$0.00371  \\
    \bottomrule
\end{tabular}
\caption{Fitting parameters for Eq.\,\ref{Eq:Fit} that empirically describes the magnon dispersion in the AFM state $(\hslash\omega_{\rm res})$ and the quasielastic line width in the AFQ state $(\Gamma_0)$. All values are given in meV.}\label{Tab:S1}
\end{center}\vspace{-2.1em}
\end{table}

For the empirical fitting of the magnon energy $(\hslash\omega_{\rm res})$ in the AFM state and the quasielastic line width $(\Gamma_0)$ in the AFQ state in 3-dimensional momentum space (Fig.\,\ref{Fig:Dispersion}), we utilized the conventional Fourier expansion, which accounts for the cubic symmetry of the Brillouin zone: $$f(\mathbf{Q}) = \sum_{j}{A_{n_j}\exp\bigl[{\mathrm{i}\,\mathbf{Q}\cdot\mathbf{R}_j} \bigr]}.$$ Here $\mathbf{Q} = (Q_x, Q_y, Q_z) = 2\pi/a(H, K, L)$ denotes a vector in the reciprocal space, the summation runs over all lattice vectors $\mathbf{R}_j = a(n_x, n_y, n_z)$, and $\,A_{n_j}$ are the free fitting parameters. The series is truncated by considering only higher-order terms with $n_j=n_x^2 + n_y^2 + n_z^2\leq16$, resulting in the following expression for the fitting function:\vspace{-0.5em}

\begin{widetext}
\begin{align} \label{Eq:Fit}
f(\mathbf{Q}) & \approx A_0 + 2 A_1 \left[ \cos(2 \pi H ) + \cos(2 \pi K) + \cos(2 \pi L)  \right] \nonumber \\
  & + 4 A_2\left[ \cos(2 \pi H )\cos(2 \pi K) + \cos(2 \pi K )\cos(2 \pi L) + \cos(2 \pi L )\cos(2 \pi H) \right] \nonumber \\
  & + 8 A_3 \left[ \cos(2 \pi H )\cos(2 \pi K)\cos(2 \pi L) \right] \nonumber \\
  & + 2 A_4 \left[ \cos(4 \pi H ) + \cos(4 \pi K) + \cos(4 \pi L)  \right] \nonumber  \\
  & + 4 A_5\left[ \cos(4 \pi H )\cos(2 \pi K) + \cos(4 \pi K )\cos(2 \pi L) + \cos(4 \pi L )\cos(2 \pi H)  \right. \nonumber \\
  & \qquad + \left. \cos(2 \pi H )\cos(4 \pi K) + \cos(2 \pi K )\cos(4 \pi L) + \cos(2 \pi L )\cos(4 \pi H) \right] \nonumber \\
  & + 8 A_6 \left[ \cos(4 \pi H )\cos(2 \pi K)\cos(2 \pi L) + \cos(2 \pi H )\cos(4 \pi K)\cos(2 \pi L) + \cos(2 \pi H )\cos(2 \pi K)\cos(4 \pi L)\right] \nonumber \\
  & + 4 A_8 \left[ \cos(4 \pi H )\cos(4 \pi K) + \cos(4 \pi K )\cos(4 \pi L) + \cos(4 \pi L )\cos(4 \pi H) \right] \nonumber \\
  & + 8 A_{9a} \left[ \cos(4 \pi H )\cos(4 \pi K)\cos(2 \pi L) + \cos(2 \pi H )\cos(4 \pi K)\cos(4 \pi L) + \cos(4 \pi H )\cos(2 \pi K)\cos(4 \pi L)\right] \nonumber \\
  & + 2 A_{9b} \left[ \cos(6 \pi H ) + \cos(6 \pi K) + \cos(6 \pi L)  \right]  \nonumber \\
  & + 4 A_{10} \left[ \cos(6 \pi H )\cos(2 \pi K) + \cos(6 \pi K )\cos(2 \pi L) + \cos(6 \pi L )\cos(2 \pi H) \right. \nonumber \\
  & \qquad + \left. \cos(2 \pi H )\cos(6 \pi K) + \cos(2 \pi K )\cos(6 \pi L) + \cos(2 \pi L )\cos(6 \pi H) \right] \nonumber \\
  & + 8 A_{11} \left[ \cos(6 \pi H )\cos(2 \pi K)\cos(2 \pi L) + \cos(2 \pi H )\cos(6 \pi K)\cos(2 \pi L) + \cos(2 \pi H )\cos(2 \pi K)\cos(6 \pi L)\right] \nonumber \\
  & + 8 A_{12} \left[ \cos(4 \pi H ) \cos(4 \pi K) \cos(4 \pi L)  \right]  \nonumber  \\
  & + 4 A_{13} \left[ \cos(6 \pi H )\cos(4 \pi K) + \cos(6 \pi K )\cos(4 \pi L) + \cos(6 \pi L )\cos(4 \pi H)  \right. \nonumber \\
  & \qquad + \left. \cos(4 \pi H )\cos(6 \pi K) + \cos(4 \pi K )\cos(6 \pi L) + \cos(4 \pi L )\cos(6 \pi H) \right] \nonumber \\
  & + 8 A_{14} \left[ \cos(6 \pi H )\cos(4 \pi K)\cos(2 \pi L) + \cos(2 \pi H)\cos(6 \pi K )\cos(4 \pi L) + \cos(4 \pi H)\cos(2 \pi K)\cos(6 \pi L ) \right. \nonumber \\
  & \qquad + \left. \cos(4 \pi H )\cos(6 \pi K)\cos(2 \pi L) + \cos(2 \pi H)\cos(4 \pi K )\cos(6 \pi L) + \cos(6 \pi H)\cos(2 \pi K)\cos(4 \pi L) \right] \nonumber \\
  & + 2 A_{16} \left[ \cos(8 \pi H ) + \cos(8 \pi K) + \cos(8 \pi L)  \right].
\end{align}
\end{widetext}

This function has been used to describe both the magnon dispersion at $T=1.5$\,K and the momentum dependence of the quasielastic line width at $T=2.6$\,K. The corresponding fitting parameters, listed in Table\,\ref{Tab:S1}, were obtained by fitting the experimental values of $\omega_{\rm res}$ and $\Gamma_0$ in the whole equatorial $(HHL)$ plane (around 530 points per temperature), which are shown in the upper half of Fig.\,\ref{Fig:Dispersion}e. The resulting fitting functions are given in the bottom part of Fig.\,\ref{Fig:Dispersion}e and as black lines in Fig.\,\ref{Fig:Dispersion}a and Fig.\,\ref{Fig:Dispersion}c.

\vfill

\end{document}